\documentclass[nohyperref]{article}

\usepackage{microtype}
\usepackage{graphicx}
\usepackage{subfigure}
\usepackage{booktabs} % for professional tables
\usepackage{url}
\usepackage{comment}
\usepackage{caption}
\usepackage{makecell}

\usepackage{hyperref}

\usepackage[accepted]{icml2022}

\usepackage{amsmath}
\usepackage{amssymb}
\usepackage{mathtools}
\usepackage{amsthm}

\usepackage[capitalize,noabbrev]{cleveref}

\theoremstyle{plain}

\theoremstyle{definition}

\theoremstyle{remark}

\usepackage[textsize=tiny]{todonotes}
\icmltitlerunning{Exploring the Effectiveness of SSL and CC in Emotion Recognition of Nonverbal Vocalizations}

\begin{document}

\twocolumn[
\icmltitle{Exploring the Effectiveness of Self-supervised Learning and Classifier Chains in Emotion Recognition of Nonverbal Vocalizations}
           
\icmlsetsymbol{equal}{*}

\begin{icmlauthorlist}
\icmlauthor{Detai Xin}{utokyo}
\icmlauthor{Shinnosuke Takamichi}{utokyo}
\icmlauthor{Hiroshi Saruwatari}{utokyo}
\end{icmlauthorlist}

\icmlaffiliation{utokyo}{The University of Tokyo, Tokyo, Japan}

\icmlcorrespondingauthor{Detai Xin}{detai\_xin@ipc.i.u-tokyo.ac.jp}

\icmlkeywords{Nonverbal Vocalizations, Emotion Recognition}

\vskip 0.3in
]


\printAffiliationsAndNotice{}  % leave blank if no need to mention equal contribution

\renewcommand{\arraystretch}{1.2}

\begin{abstract}
We present an emotion recognition system for nonverbal vocalizations (NVs) submitted to the ExVo Few-Shot track of the ICML Expressive Vocalizations Competition 2022.
The proposed method uses self-supervised learning (SSL) models to extract features from NVs and uses a classifier chain to model the label dependency between emotions.
Experimental results demonstrate that the proposed method can significantly improve the performance of this task compared to several baseline methods.
Our proposed method obtained a mean concordance correlation coefficient (CCC) of $0.725$ in the validation set and $0.739$ in the test set, while the best baseline method only obtained $0.554$ in the validation set.
We publicate our code at \url{https://github.com/Aria-K-Alethia/ExVo} to help others to reproduce our experimental results.
\end{abstract}
\vspace{-10mm}
\section{Introduction}
\vspace{-1mm}
Nonverbal vocalizations (NVs), also called affect bursts, refer to short and expressive vocalizations containing no linguistic information like laughter, sobs, and screams~\cite{scherer1994affect, trouvain2012comparing}.
%They are relatively casual expressions and are usually not used in written languages~\cite{trouvain2012comparing}.
NVs are important for spoken language processing, since (1) NVs play an important role in expressing emotions in spoken languages~\cite{hall2009psychosocial, scherer2011assessing}, and (2) they are common components of human communication existing in different cultures and languages~\cite{sauter2010cross}.
Although both emotional prosody and NVs contribute to emotional expressions in speech, NVs are ignored by most previous research on speech emotions~\cite{lima2013voices}, which necessitates further work in this field.

The ICML Expressive Vocalizations Competition (hereafter ExVo) launched in 2022 aims to develop technologies for the recognition, generation, and personalization of NVs~\cite{baird2022icml}.
The competition includes three tracks: ExVo Multi-Task, ExVo Generate, and ExVo Few-Shot, which correspond to the three goals of recognizing, generating, and personalizing NVs, respectively.
Specifically, the ExVo Multi-Task track aims to recognize not only emotions but also demographic information like the age and native country of the speakers from NVs.
In the ExVo Generate track, the participants need to generate NVs of various emotions.
Similar to the Multi-Task track, the ExVo Few-Shot track also aims to recognize emotions from NVs, but further requires the prediction systems to adapt to new speakers which are not in the training set.
All tracks are evaluated by appropriate subjective and objective metrics to reflect the performance of the proposed systems.

In this paper, we describe our emotion recognition system for NVs submitted to the ExVo Few-Shot track.
Our system consists of two components: a feature extractor and a classifier chain (CC).
As the feature extractor, we use a self-supervised learning (SSL) model like Wav2vec2~\cite{schneider2019wav2vec, baevski2020wav2vec} or HuBERT~\cite{hsu2021hubert}.
The extracted feature is then fed to a classifier chain, which predicts the score for each emotion sequentially.
The proposed classifier chain predicts the score by conditioning on both the extracted feature and the predicted scores in previous steps, so can utilize information from label dependency.
We conducted comprehensive experiments to verify the effectiveness of these two components, and show that the proposed method can significantly improve the performance compared to several baseline methods.
Our contributions can be summarized as follows:
\vspace{-3mm}
\begin{itemize} \itemsep -1mm 
    \item We propose an emotion recognition system for NVs using SSL models and CC, and conduct experiments to show the effectiveness of the proposed method.
    \item We conduct experiments to show the best SSL models for this task.
    \item We analyze the prediction results and give insights for future research.
\end{itemize}

\vspace{-6mm}
\section{The ExVo Few-Shot track}
\vspace{-1mm}
The task of the ExVo Few-Shot track is recognizing emotions from NVs for unseen speakers that are not in the training set.
During the training phase, the participants can train their emotion recognition models on the data of the seen speakers.
In the test phase, two samples for each unseen speaker will be provided by the organizer, and the participants can then adapt their models to unseen speakers by few-shot learning.

ExVo provides a large-scale multilingual (Chinese, English, Spanish) NVs dataset~\cite{Cowen2022HumeVB}, which contains approximately $36$ hours NVs uttered by $1,702$ speakers from the USA, China, South Africa, and Venezuela.
Ten emotions consisting of amusement, awe, awkwardness, distress, excitement, fear, horror, sadness, surprise, and triumph are covered by the dataset.
Each NV is annotated with ten emotion scores ranging from $0$ to $1$ by crowd-sourcing.
Some statistics of the dataset are summarized in Table~\ref{table:dataset}.
\begin{table}[t]
\caption{Statistics of the dataset. Some terms are not available for the participants.}
\label{table:dataset}
\begin{center}
\begin{small}
%\begin{sc}
\begin{tabular}{l|ccc|c}
\toprule
            & Train & Validation & Test  & $\sum$\\
\midrule
Duration (hrs) & $12.32$ & $12.10$ & $12.37$ & $36.78$ \\
Speakers    &  $571$   & $568$ &  $563$ & $1702$    \\
Audio clips & $19,990$ &  $19,396$ & $19,815$  &   $59201$  \\
\makecell[ll]{Audio clips\\\;\;per speaker} & \makecell{$4$-$201$\\(Avg. $35$)} & \makecell{$4$-$168$\\(Avg. $34$)} & --  &  -- \\
\bottomrule
\end{tabular}
\vspace{-8mm}
%\end{sc}
\end{small}
\end{center}
\end{table}
The participants should submit the predictions of their models on test samples.
Concordance correlation coefficient (CCC) is used to evaluate the performance, which can measure the agreement between the ground-truth (GT) and predicted emotion scores.

\vspace{-3mm}
\section{Proposed method}
\vspace{-1mm}
In this section, we describe the proposed method.
We first introduce the general architecture of the proposed method and then describe each component separately.

\vspace{-4mm}
\subsection{Architecture}
\begin{figure}[t]
\begin{center}
\centerline{
\includegraphics[width=0.7\columnwidth]{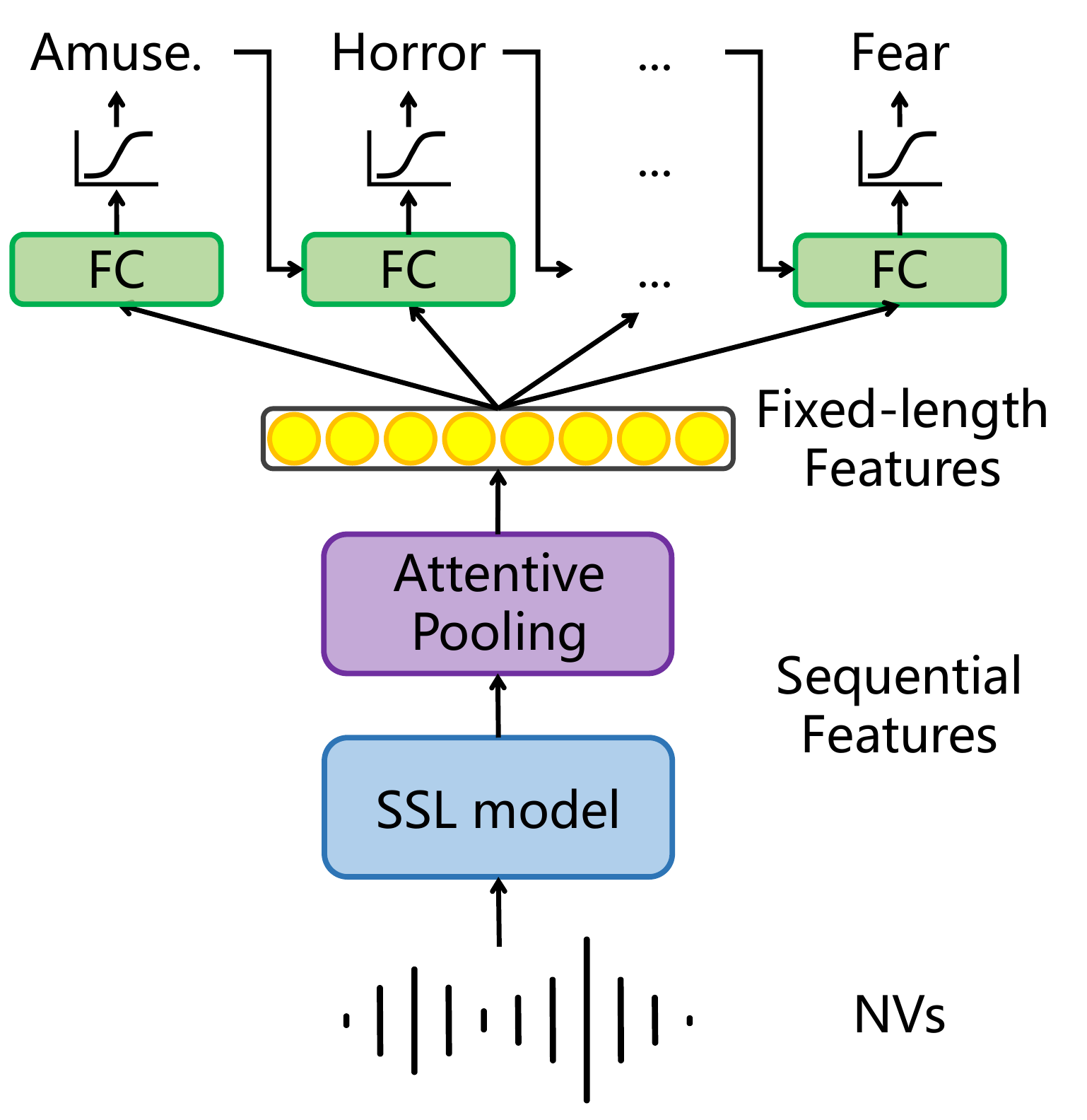}
}
\caption{Architecture of the proposed method. FC denotes a fully connected layer.}
\label{figure:achitecture}
\end{center}
\vspace{-12mm}
\end{figure}

\vspace{-1mm}
The architecture of the proposed method is illustrated in Figure~\ref{figure:achitecture}.
Specifically, the NVs are first fed to the SSL model to extract sequential features from them.
Then, an attentive pooling module is used to collect information over the time axis and convert the sequential features into fixed-length features.
Finally, a CC is used to predict the score for each emotion sequentially.
Each emotion has a separate predictor, which contains a fully connected (FC) layer followed by a sigmoid activation.
The predictor in the CC not only uses the extracted features but also the predicted scores of previous classifiers as input, which can thus utilize information of label dependency.

\vspace{-4mm}
\subsection{Self-supervised learning models}
\vspace{-1mm}
SSL is a popular and powerful method to leverage large-scale unlabeled data.
The idea of SSL is to set a sophisticated pretext task to learn nontrivial data representations.
Recently, several works based on SSL have been proposed in speech processing, which showed promising results in speech-to-text and other speech-based tasks~\cite{schneider2019wav2vec, baevski2020wav2vec, conneau2020unsupervised, hsu2021robust, hsu2021hubert}.
Although previous work has shown that pretrained SSL models were effective on speech emotion recognition in verbal communication\cite{pepino2021emotion,chen2021exploring, wang2021fine, wagner2022dawn}, since most of the SSL models were trained on speech corpora that rarely contain NVs, whether these models are effective on NVs remains unknown.
Therefore, in the experiments, we select several SSL models and verify their effectiveness in the task of recognizing emotions from NVs.

\vspace{-4mm}
\subsection{Classifier chain}
\vspace{-1mm}
The task of the ExVo Few-Shot track is a multivariate regression problem.
While similar tasks like continuous speech emotion recognition usually use separate predictors to predict the emotions independently~\cite{atmaja2021evaluation}, it is noteworthy that the emotion labels in the provided dataset have an intrinsic dependency.
For example, if the score of amusement is high, the scores of negative emotions like sadness and fear may be low, which implies the possibility to utilize information from label dependency.
To this end, we propose to use CC~\cite{read2011classifier}, which has been proved to be useful for modeling label dependency in multi-label classification~\cite{dembszynski2010label, dembczynski2010bayes, nam2017maximizing}.

Formally, denoting the predictor of the $i$-th emotion and the extracted feature as $f_{i}(\cdot)$ and $z$, respectively, the emotion score $\widehat{y}_{i}$ is computed by:
$
    \widehat{y}_{i} = \sigma (f_{i}(z \oplus \widehat{y}_{<i})),
$
where $\sigma(\cdot)$ is the sigmoid function, $\oplus$ is the vector concatenation operator, and $\widehat{y}_{<i}$ indicates a vector concatenating previously predicted emotion scores.
During training, we feed GT emotion scores $y_{<i}$ to the predictor to prevent error propagation.

While it is possible to use powerful sequential models like a recurrent neural network to implement the predictor $f_{i}$ in CC~\cite{nam2017maximizing}, in the preliminary experiments we found simple linear CC was more stable than other models, thus we use linear CC in the proposed method.
Chain order is another critical problem for CC since it determines what information can be utilized by each classifier~\cite{read2021classifier}.
In our implementation, we adopt a heuristic method in which we first train ten separate base predictors for each emotion, then use the descending order of the performance of the base predictors as the chain order.
This method can intuitively alleviate the error propagation problem by first predicting emotions with high confidence, hence can improve the performance~\cite{read2021classifier}.

\vspace{-4mm}
\subsection{Loss function}
\vspace{-1mm}
We use CCC as the objective for the model training.
Previous work has demonstrated that CCC was better for speech emotion recognition than error-based loss functions like L1 loss~\cite{atmaja2021evaluation}.
Generally, given a set of paired data ($\{s^{j}\}_{j=1}^{n}, \{t^{j}\}_{j=1}^{n}$) with length $n$, the CCC between them is defined as:
$
    \mathrm{CCC}(\{s^{j}\}_{j=1}^{n}, \{t^{j}\}_{j=1}^{n}) = \frac{2\sigma^{2}_{st}}{\sigma^{2}_{s} + \sigma^{2}_{t} + (\mu_{s} - \mu_{t})^{2}}.
$
In the proposed method the objective value of a mini-batch is computed by averaging CCC values of all emotions.
Formally, denoting the number of emotions and the mini-batch size as $C$ and $B$, respectively, the loss function of the proposed model is defined as:
$
    \mathcal{L} = \frac{1}{C}\sum_{i=1}^{C} \mathrm{CCC}(\{y_{i}^{j}\}_{j=1}^{B}, \{\widehat{y}_{i}^{j}\}_{j=1}^{B}),
$
where $y_{i}^{j}$ and $\widehat{y}_{i}^{j}$ are GT and predicted scores of the $i$-th emotion of the $j$-th sample in the mini-batch, respectively.

\vspace{-4mm}
\subsection{Data augmentation}
\vspace{-1mm}
To improve the robustness of the model, we additionally use data augmentation.
Specifically, we use two strategies: pitch-shifting and speaking-rate-changing~\cite{saeki2022utmos}.
Pitch-shifting raises or lowers the pitch of NVs, and can change the speaker identity of NVs.
Speaking-rate-changing slows down or speeds up the NVs.
We tune the shifting range and the speaking-rate-changing range so that the emotion of the augmented NVs has little difference from the original ones.
\vspace{-1mm}
\section{Experiments}
\begin{table}[t]
\caption{Average CCC of different SSL models in $5$-fold cross-validation. \textbf{Bold} indicates the best score.}
\vspace{1mm}
\label{table:ssl}
\begin{center}
\begin{small}
%\begin{sc}
\begin{tabular}{lc}
\toprule
SSL Model & CCC \\
\midrule
Wav2vec2-base   &  $0.699 \pm 0.013$ \\
HuBERT-base     &  $0.703 \pm 0.013$ \\
Wav2vec2-robust &  $0.709 \pm 0.012$ \\
Wav2vec2-large  &  $0.709 \pm 0.009$ \\
HuBERT-large    &  $0.718 \pm 0.011$ \\
XLSR            &  $\mathbf{0.722 \pm 0.012}$ \\
\bottomrule
\end{tabular}
%\end{sc}
\end{small}
\end{center}
\vspace{-6mm}
\end{table}
\begin{table}[t]
\caption{Average CCC of the proposed method using CC and data augmentation in 5-fold cross-validation. \textbf{Bold} indicates the best score.}
\vspace{1mm}
\label{table:ablation}
\begin{center}
\begin{small}
%\begin{sc}
\begin{tabular}{lc}
\toprule
Model & CCC \\
\midrule
CC HuBERT-large      &  $0.722 \pm 0.012$ \\
CC HuBERT-large Aug.     &  $0.722 \pm 0.012$ \\
CC XLSR              &  $0.724 \pm 0.012$ \\
CC XLSR Aug.         &  $\mathbf{0.726 \pm 0.013}$ \\
\bottomrule
\end{tabular}
%\end{sc}
\end{small}
\end{center}
\vspace{-6mm}
\end{table}
\begin{table*}[t]
\setlength{\tabcolsep}{1.5mm}
\caption{CCC of the proposed and baseline methods for each emotion on all validation data. \textbf{Bold} indicates the best score with p-value $< 0.05$.}
\vspace{1mm}
\label{table:final}
\begin{small}
%\begin{sc}
\begin{tabular}{l|cccccccccc|c}
\toprule
Model & Awe & Excite. & Amuse. & Awkward. & Fear & Horror & Distress & Triumph & Sadness & Surprise & Average \\
\midrule
DeepSpectrum & $0.575$ & $0.424$ & $0.512$ & $0.318$ & $0.497$ & $0.480$ & $0.356$ & $0.263$ & $0.412$ & $0.621$ & $0.446$ \\
eGeMAPS & $0.592$ & $0.420$ & $0.525$ & $0.302$ & $0.527$ & $0.511$ & $0.365$ & $0.263$ & $0.407$ & $0.604$ & $0.452$ \\
BoAW   & $0.639$ & $0.470$ & $0.583$ & $0.383$ & $0.566$ & $0.548$ & $0.457$ & $0.347$ & $0.472$ & $0.659$ & $0.512$ \\
ComParE     & $0.680$ & $0.503$ & $0.638$ & $0.403$ & $0.604$ & $0.587$ & $0.494$ & $0.411$ & $0.526$ & $0.704$ & $0.554$ \\
\midrule
CC XLSR Aug. & $\mathbf{0.801}$ & $\mathbf{0.687}$ & $\mathbf{0.796}$ & $\mathbf{0.608}$ & $\mathbf{0.755}$ & $\mathbf{0.735}$ & $\mathbf{0.685}$ & $\mathbf{0.674}$ & $\mathbf{0.703}$ & $\mathbf{0.804}$ & $\mathbf{0.725}$ \\
\bottomrule
\end{tabular}
%\end{sc}
\end{small}
\vskip -0.1in
\end{table*}
\vspace{-1mm}
\subsection{Experimental Setup}
\vspace{-1mm}
We constructed several baseline systems based on the features provided by the organizer~\cite{baird2022icml}.
These features include: the $6373$-dimensional ComParE set from the 2016 Computational Paralinguistics Challenge~\cite{schuller2016interspeech} and the $88$-dimensional extended Geneva Minimalistic Acoustic Parameter Set (eGeMAPS)~\cite{eyben2015geneva} extracted by openSMILE toolkit~\cite{eyben2010opensmile}, the Bag-of-Audio-Words (BoAW) representations with a $2000$ codebook size extracted by openXBOW toolkit~\cite{schmitt2017openxbow} from the low-level descriptors of ComParE set, the $4096$-dimensional spectrum representations extracted by the DeepSpectrum toolkit~\cite{amiriparian2017snore}.
We used multi-layer neural networks with LeakyReLU activation~\cite{Maas13rectifiernonlinearities} to process each of the features.
Batch normalization~\cite{ioffe2015batch} was used in each layer to normalize the features.

To verify the effectiveness of SSL models and find the best model for NVs, we selected six SSL models: Wav2vec2-\{base, large\}~\cite{baevski2020wav2vec}, HuBERT-\{base, large\}~\cite{hsu2021hubert}, Wav2vec2-robust~\cite{hsu2021robust}, and XLSR~\cite{conneau2020unsupervised}.
Wav2vec2-\{base, large\} and HuBERT-\{base, large\} are common models used in previous work~\cite{chen2021exploring, wang2021fine}.
The Wav2vec2-\{base, large\} and HuBERT-large were trained on the Librispeech~\cite{panayotov2015librispeech} corpus containing $960$ hours of audio, while the HuBERT-large model was trained on the Libri-Light~\cite{2020librilight} corpus containing $60$k hours of audio.
These corpora are all from clean audiobooks that rarely contain NVs.
Also, since the NVs in the dataset were recorded by the speakers themselves and sometimes contain noises, we selected Wav2vec2-robust that was trained on not only the Libri-Light corpus but also noisy telephone corpora including Switchboard~\cite{godfrey1997switchboard}, Fisher~\cite{cieri2004fisher}, and CommonVoice~\cite{ardila2019common}.
Finally, to handle multilingual NVs we select XLSR, which is a multilingual version of Wav2vec2 trained on $56$k hours of multilingual audio (multilingual Librispeech~\cite{pratap2020mls}, CommonVoice, and Babel~\cite{gales2014speech}) covering $53$ languages.

Instead of using the original train-validation split, we first combined the train and validation sets and then used $5$-fold cross-validation to train all models.
To ensure the validation results can reflect the generalization ability of the models, we split the data by speaker so that the speakers of the validation set of each split are unseen for the training set.
Each split had about $912$ seen speakers and $227$ unseen speakers.
All silence in the audios was trimmed by voice activity detection.
The chain order was set to awe, surprise, amusement, fear, horror, sadness, distress, excitement, triumph, and awkwardness.
We used Adam~\cite{kingma2014adam} to optimize the model, the learning rate was set to $1\mathrm{e}{-4}$ for the CC and $1\mathrm{e}{-5}$ for the feature extractor.
The learning rate was halved once the loss value on the validation set did not decrease.
The batch size was set to $16$ when Wav2vec-base or HuBERT-base was used, otherwise, it was set to $8$.
We used WavAugment~\cite{wavaugment2020} to implement the data augmentation, the pitch-shifting range was set to $[-300, 300]$ cents, and the speaking-rate-changing range was set to $[0.8, 1.2]$.
The minimal and maximal number of epochs were set to $10$ and $50$, respectively.
Early stopping was used to prevent overfitting with a patience of $10$ epoch.
We selected the model that had the best CCC on the validation set.
To adapt the models to unseen speakers, we further fine-tuned the models on the $2$ samples of each test speaker, which roughly took $10$ epochs for convergence.
The learning rate was set to $1\mathrm{e}{-6}$ for the whole network.

\vspace{-1mm}
\subsection{Evaluations of SSL models}
\vspace{-1mm}
We first evaluate all SSL models to verify their effectiveness and find the best model for NVs.
To avoid the the influence on performance from the CC, in this experiment the CC is replaced by a single FC layer. 
The results are shown in Table~\ref{table:ssl}.
It can be seen that the XLSR model obtained the best performance.
We suppose this is because the multilingual XLSR model can handle more phonetic tokens of different languages, unlike other SSL models that are only trained on English corpora.
Also, large models are always better than base models, which is consistent with the results of previous work~\cite{wagner2022dawn}.

\vspace{-1mm}
\subsection{Evaluations of CC and data augmentation}
\vspace{-1mm}
We select the top-2 SSL models (XLSR and HuBERT-large) in the previous experiment as the feature extractors to evaluate CC and data augmentation (``Aug.'').
The results are shown in Table~\ref{table:ablation}.
It can be seen that CC consistently improves the performance compared to the results in Table~\ref{table:ssl}
The data augmentation brings improvements for the XLSR model but failed to improve the performance of the CC HuBERT-large model.

\vspace{-1mm}
\subsection{Evaluations for each emotion}
\vspace{-1mm}
We then evaluate the proposed method for each emotion.
We selected the best performing model CC XLSR Aug. obtained in previous experiments.
All baseline models were trained using cross-validation for comparison.
After training, we combined the inference results on all $5$ splits of each model and computed the CCC value for each emotion.
The result is demonstrated in Table~\ref{table:final}.
It can be seen that the proposed method significantly outperformed all baseline methods on all emotions, which indicates the effectiveness of SSL models and CC.
Besides, it can be observed that all models have difficulties to recognize some emotions like awkwardness, triumph, distress.

\vspace{-1mm}
\subsection{Evaluations on test set}
\vspace{-1mm}
We finally fine-tuned the best model (CC XLSR Aug.) on the test samples provided by the organizer.
Our best CCC on the test set is $0.739$, which demonstrates the proposed method has a strong generalization ability.
\vspace{-3mm}
\section{Conclusions}
\vspace{-1mm}
This paper described an emotion recognition system for NVs submitted to the Few-Shot track of the ICML ExVo competition.
The proposed method uses a SSL model to extract contextual speech representations from NVs, and uses a CC to predict emotion scores by utilizing the features and the emotion scores predicted in previous steps together.
Experimental results demonstrated that the proposed method significantly outperformed several baseline methods.

{\normalsize
\textbf{Acknowledgements:} 
This work was supported by JST SPRING, Grant Number JPMJSP2108. Part of this work was also supported by JSPS KAKENHI Grant Number 21H04900 (for implementation) and JST Moonshot R\&D Grant Number JPMJPS2011 (for evaluation).
}

\bibliography{mybib}
\bibliographystyle{icml2022}
\end{document}